\documentclass[11pt]{article}  %Do not change the point size.
\usepackage{menuproc}
\usepackage{epsfig}
\def\be{\begin{equation}} 
\def\ee{\end{equation}} 
\def\bea{\begin{eqnarray}} 
\def\eea{\end{eqnarray}}

\begin{document}
%
%____________________________________________________________
%
%  Title, authors, institutions, and abstract
%----------------------------------------------------------------
%  Syntax:  \titlematter{title}{authors}{institutions}{abstract}
%----------------------------------------------------------------
%     If lines are too long, use linebreaks where convenient.
%     If all authors are from the same institution, omit raised letters.
%
\titlematter{Muon capture in hydrogen}
{S. Ando, F. Myhrer and K. Kubodera}%
{Department of Physics \& Astronomy,\\
 University of South Carolina, Columbia, SC 29208, U.S.A. }%
{
%The abstract should give a brief description of the article's content, not
%exceeding 10 lines for plenary papers and 5 lines for parallel contributions.
%---
Theoretical difficulties in reconciling the 
measured rates for 
ordinary and radiative muon capture are discussed, 
based on heavy-baryon chiral perturbation theory. 
We also examine ambiguity in our analysis due to
the formation of $p\mu p$ molecules 
in the liquid hydrogen target. 
}
% 
%--------------------------------------------------------------------------
%\begin{figure}[b]
%\centerline{\epsfig{file= pin_news.eps,width=.4\textwidth,silent=,clip=}}
%\centerline{\bfseries No. 16, December 2001}
%\caption{\label{pin}
%The figures may be set centered, as in this figure.}
%\end{figure}
%--------------------------------------------------------------------------- 
%  Start article here: 
%%%%%%%%%%%%%%%%%%%%%%%%%%%%%%%%%%%%%%%%%%%%%%%%%%%%%%%%%%%%%%%%%%%%%%%%%%
\section{Introduction}
%The Proceedings of the MENU2001 Conference will be dedicated
%to Professor Gerhard H{\"o}hler, one of the founding fathers of the MENU 
%conference
%series, on the occasion of the 80th anniversary of his birthday.
%%%%%%%%%%%%%%%%%%%%%%%%%%%%%%%%%%%%%%%%%%%%%%%%%%%%%%%%%%%%%%%%%%%%%%%%%%%%%%%%

Ordinary and radiative muon captures 
(OMC and RMC) on a proton,  
$\mu^-p \to n\nu_\mu $ and 
$\mu^-p \to n\nu_\mu\gamma $,  
are fundamental weak-interaction processes 
in nuclear physics 
and constitute primary sources of information on $g_P$, 
the induced pseudoscalar coupling constant 
of the weak nucleon current \cite{morita}. 
The most accurate existing measurements 
of the OMC and RMC rates have been carried out
using a liquid hydrogen target. 
The experimental OMC rate
in liquid hydrogen obtained by 
Bardin {\it et al.}~\cite{bardin} is
\bea
\Lambda_{liq}^{exp} = 460 \pm 20 
\ \ \ \mbox{\rm [s$^{-1}$]}   
\ \ \ \ \ \ {\rm (OMC)} .
\label{eq;exp-omc}
\eea
As for RMC, Jonkmans {\it et al.}~\cite{jonkmans96} 
measured the absolute photon spectrum for 
$E_\gamma\ge 60$ MeV and deduced therefrom
the partial RMC branching ratio, $R_\gamma$, 
which is the number of RMC events (per stopped muon)
producing a photon with $E_\gamma\ge 60$ MeV.
The measured value of $R_\gamma$ is \cite{jonkmans96,wright98}
\bea
R^{exp}_\gamma = (2.10\pm 0.22)\times 10^{-8}
\ \ \ {\rm (RMC)}  .
\label{eq;exp-rmc}
\eea
Surprisingly, the value of $g_P$
deduced in \cite{jonkmans96,wright98}
from the RMC data
is $\sim$1.5 times larger than the PCAC prediction, 
$g_P^{PCAC}$ \cite{PCAC}.
By contrast, the value of $g_P$ deduced in \cite{bardin}
from the OMC data is in good agreement with 
the PCAC prediction. 
Heavy-baryon chiral perturbation theory (HBChPT),
a low-energy effective theory of QCD, 
allows us to go beyond the PCAC approach,
but the results of detailed HBChPT calculations  
up to next-to-next-to-leading order ~\cite{bkm94}
essentially agree with those obtained in the PCAC approach.
Thus the theoretical framework for 
estimating $g_P$ appears to be robust. 
What then can be the origin of the apparent conflict 
between the $g_P$ values determined from OMC and RMC ? 
In this talk we wish to address a number of issues
relevant to this question.

The OMC process is described by the 
standard electroweak theory.  
The lepton current, 
$j_\alpha^{lepton}$, interact with the very heavy 
$W$ boson ($m_W \simeq $ 80 GeV)  
which propagates  
and interacts with the hadronic current, 
$J_\beta^{hadron}$. 
%The axial component of the $W$ propagator could be extended as  
%a pion propagator which subsequently interacts 
%with the nucleon. 
%It is this so-called ``pion-pole" which 
%generates the 
%induced pseudoscalar coupling. 
Since the maximal momentum transfer 
between the two currents, 
$q_{max} \simeq m_\mu \ll m_W$,  
we can for all practical purposes replace the 
$W$ boson propagator 
with a constant.   
Then the interaction between the lepton and hadron currents 
reduces to:
\be
\sim \frac{G_F}{\sqrt{2}} j_\alpha^{lepton}
 J_\alpha^{hadron} . 
\ee 
where $G_F$ is the effective Fermi constant. 
Since %(as far as we know) 
the leptons are considered point particles,  
the lepton current is given 
by the incoming muon and outgoing neutrino spinors: 
%\be
$ j_\alpha^{lepton} = \bar u_\nu \gamma_\alpha 
(1-\gamma_5) u_\mu . $
%\ee 
The hadronic current however is less well known 
due to the structure of the nucleons. 
We write $J_\beta^{hadron} = V_\beta - A_\beta$\,,  
where the vector current, $V_\beta$, and 
the axial current, $A_\beta$, 
can be written as 
(based on symmetry considerations and  
neglecting second-class currents): 
\be
V_\beta = \bar u_n \left[ g_V(q^2) \gamma_\beta + 
i \; \frac{ g_M(q^2) }{2 m_N} \; 
\sigma_{\beta \delta} q^\delta \right] u_p , 
\label{eq;vector}
\ee
\be 
A_\beta = \bar u_n \left[ g_A(q^2) \gamma_\beta \gamma_5 + 
\frac{g_P(q^2)}{m_\mu} \; q_\beta \gamma_5 \right] u_p\, , 
\label{eq;axial}
\ee 
where $u_n$ and $u_p$ are the outgoing neutron 
and incoming proton spinors; and 
$q$ is the momentum transferred to the nucleon.
The four form factors, 
$g_V(q^2)$, $g_M(q^2)$, $g_A(q^2)$ and $g_P(q^2)$, 
in Eqs.(\ref{eq;vector}) and (\ref{eq;axial}) 
are, at low momentum transfers relevant to the 
OMC and RMC reactions,  
given by known parameters: 
\bea 
g_V(q^2) &=& 1 + \frac{1}{6} <\!r^2\!> q^2 + \cdots \ ; 
\; \; \; \; \; \; \; \; \; \; 
g_M(q^2) = \kappa_p - \kappa_n + \cdots \ ;  \\ 
g_A(q^2) &=& g_A \left( 1 + \frac{1}{6} <r^2_A > q^2 + 
\cdots \right) \ ; \\ 
g_P(q^2) &=& \frac{2 m_\mu f_\pi g_{\pi NN}}{m_\pi^2 - q^2} 
- \frac{1}{3} g_A m_\mu m_N <\!r^2_A\!> + \cdots \ \ . 
\label{pseudoscalar}
\eea 
Here $<\!r^2\!>$ $\simeq 0.585$ fm$^2$ %\cite{rV}
is the square of the 
isovector nucleon radius, 
$<\!r^2_A\!>$ $\simeq $ 0.42 fm$^2$ %\cite{rA}
is the axial radius squared, and 
$\kappa_p$ and  $\kappa_n$ are the proton and 
neutron anomalous magnetic moments, respectively. 
In the induced pseudoscalar form factor, 
$ g_P(q^2)$, 
the form of the dominant ``pion-pole" term 
is dictated by the PCAC hypothesis. 

In OMC the momentum transfer is 
$q^2$=$(m_\mu -E_\nu)^2 - E_\nu^2$
=$- 0.88 m_\mu^2 $ and, traditionally,
the pseudoscalar coupling ``constant", $g_P$,
is defined as 
$g_P \equiv g_P(q^2\!=\!- 0.88 m_\mu^2)$,
and thus the PCAC value is 
$g_P^{PCAC}=6.77g_A(0)$~\cite{fea80}. 
We remark that the contribution 
of the $g_P(q^2)$ term to OMC
is suppressed due to the fact 
that  $q^2$=$- 0.88 m_\mu^2$ is far from the pion pole.
The rationale for using RMC
to determine $g_P$ 
is that the three-body kinematics 
in the final state allows $q^2$ to approach the pion pole,
enhancing thereby the contribution
of the pseudoscalar term. 
With the photon energy denoted by $E_\gamma$,
we have $q^2$ $\simeq (2 m_\mu E_\gamma - m_\mu^2)$, 
which is positive for sufficiently large values 
of $E_\gamma$, and which can reach the 
maximal value $\simeq m_\mu^2$.
In \cite{jonkmans96,wright98},
the RMC process is measured for 
$E_\gamma$ $>$ 60 MeV. 
%
%can be understood by rewriting 
%Eq.~(\ref{pseudoscalar}) as
%\bea
%g_P(q^2) \simeq g_P \, 
%\left(\frac{m_\pi^2+0.88m_\mu^2}{m_\pi^2 - q^2}\right) . 
%\label{GP}
%\eea
Meanwhile, a great challenge in observing RMC 
is its very low branching ratio 
($\sim 10^{-3}$ compared to OMC),
and it was quite a feat that the 
TRIUMF group succeeded in measuring $R_\gamma$.  
It should also be mentioned
that the absolute photon spectrum 
(for $E_\gamma \ge$ 60 MeV) 
measured in \cite{jonkmans96,wright98} 
gives information on the $q^2$ dependence of $g_P(q^2)$. 
The afore-mentioned surprising conclusion 
that, to fit the TRIUMF RMC data, 
one needs to adopt $g_P \sim 1.5 \; g_P^{PCAC}$ 
has triggered re-examination of the 
foundation of the theoretical framework
used for the analysis. 
In particular, it motivated detailed
systematic calculations based on HBChPT
for RMC as well as OMC 
\cite{mmk98,am98,amk01,bhm01,fearing-omc}. 

\section{Chiral perturbation theory and the atomic capture rates}

Chiral perturbation theory (ChPT) 
is an effective field theory of QCD 
tailored for describing low-energy hadronic interactions. 
Since the physical degrees of freedom
here are hadrons, not quarks or gluons, 
we ``integrate out" the quark and gluon fields 
to obtain an effective lagrangian, ${\cal L}_{eff}$, 
pertinent to  the hadronic degrees of freedom. 
${\cal L}_{eff}$ should inherit 
all the symmetry properties of QCD
(and the patterns of their breaking, if any),
including chiral symmetry. 
For our purposes,
it is sufficient to consider only 
two flavors ($u$ and $d$);
correspondingly, ${\cal L}_{eff}$ 
contains only the nucleon and pion fields. 
If chiral symmetry is an exact symmetry of ${\cal L}_{eff}$, 
then the right- and left-handed nucleon fields decouple, 
and the left- and right-handed Noether currents 
are separately conserved:
$\partial_\alpha \; j_{left}^\alpha$ = 0 and 
$\partial_\alpha \; j_{right}^\alpha$ = 0. 
The QCD vacuum state, however,
does not respect chiral symmetry, i.e. 
chiral symmetry is spontaneously broken. 
Then, according to the Goldstone theorem,
there appear massless pseudoscalar Goldstone fields,
which can be identified with (massless) pions. 
In reality,  ${\cal L}_{QCD}$ 
contains a tiny quark mass term
($m_{quark} \simeq  $10 MeV)
that violates chiral symmetry
and that gives rise to a finite pion mass. 
A consequence of the non-zero mass is that 
the axial current is no longer conserved. 
Since $m_{quark} \ll \Lambda_{chiral} \simeq $ 1 GeV, 
this explicit chiral symmetry breaking
can be treated as a small perturbation, 
and ChPT consists in expanding
${\cal L}_{eff}$ and transition amplitudes
in terms of $Q/\Lambda_{chiral}$
and $m_{pion}/\Lambda_{chiral}$,
where $Q$ is a typical scale of external momenta:
$Q \sim |\vec{p}_{pion}| \sim
|\vec{p}_{nucleon}|$. 
A version of ChPT that is particularly useful
for our present purposes is HBChPT,
wherein the nucleon is treated as a heavy particle
for which a Foldy-Wouthuysen-like 
non-relativistic expansion can be used.
%To lowest order in this expansion, 
%we recover the GOR relation  
%%\be
%$ m_\pi^2 = -2 \frac{m_{quark}}{f_\pi^2}  
%\langle 0 | \bar q q | 0 \rangle .  $
%\nonumber
%\ee 
In HBChPT we have an expansion 
\bea
{\cal L}_{eff} 
%&\sim & \sum_{\beta = 1} {\cal A}_\beta \; 
%\left(\frac{Q}{\Lambda_{chiral}} \right)^\beta 
% = {\cal A}_1 \left(\frac{Q}{\Lambda_{chiral}} \right) + 
%{\cal A}_2 \left(\frac{Q}{\Lambda_{chiral}} \right)^2 
%+ \cdots \nonumber \\ 
&=& \sum {\cal L}_\beta =  
{\cal L}_1 +  {\cal L}_2 +\cdots . 
\eea 
Here the ``chiral index" $\beta$ is given by 
$\beta = d+n/2-1$, 
where $n$ is the number of nucleon fields 
involved in a given vertex
and $d$ the number of derivatives or powers 
of $m_\pi$ involved.
For example, the lowest chiral order lagrangian is
given as  
%(below $v$ is the nucleon four-velocity 
%$= (1,\vec{0})$, i.e. only the time component is non-zero). 
\bea
{\cal L}_1 &=& \bar{N}
\left[i \; \frac{ \partial}{\partial t} +  
 \frac{g_A}{2 f_\pi} \; \mbox{\boldmath $\tau$} 
\cdot ( \vec{\sigma}\cdot 
\vec \nabla \mbox{\boldmath $\pi$}) \right]N 
+\frac{1}{2}\left(\partial_\beta \mbox{\boldmath $\pi$} \cdot 
\partial^\beta \mbox{\boldmath $\pi$} \right) 
- \frac{1}{2} m_\pi^2 \mbox{\boldmath $\pi$}^2 +\cdots , 
%+f_\pi^2{\rm Tr}\left(-\Delta\cdot\Delta+
%\frac{\chi_+}{4}\right),
\eea
while the next order Lagrangian is given by 
\bea
{\cal L}_2 &=& 
\bar{N}\left[ \frac{\vec \nabla ^2}{2 m_N} + \cdots \right] N 
+ \cdots 
%\frac{1}{2m_N}\bar{N}\left[(v\cdot D)^2-D^2+
%2g_A\{v\cdot \Delta,S\cdot D\}
%-i(1+b_5)[S^\alpha,S^\beta]f^+_{\alpha\beta}
%\right]N ,
%\\
%{\cal L}_2 &=& \frac{1}{(4\pi f_\pi)^2}\bar{N}\left[
%c_3v^\alpha [D^\beta,f^+_{\alpha\beta}]
%+c_{13}g_AS^\alpha[D^\beta,f^-_{\alpha\beta}]
%+ic_{14}g_AS^\alpha[D_\alpha,\chi_-]
%\right] N+{\cal L}_{1/m_N^2} ,
\eea
It is to be noted that in HBChPT 
the Schroedinger operator $\vec \nabla ^2 / (2 m_N)$
for the nucleon kinetic energy is treated
as a ``recoil" correction to ${\cal L}_1$. 
In muon capture (both OMC and RMC), 
$Q\sim m_\mu$=105.7 MeV 
or $Q/\Lambda_\chi\sim 0.1$ 
and hence the chiral expansion is expected
to converge rapidly.
The explicit calculations in HBChPT
\cite{am98,fearing-omc} %,fearing-omc-rmc,FS}
corroborates this expectation. 

For OMC, leading-order (LO) contributions in chiral counting
come from two tree diagrams 
of order $(Q / \Lambda_{chiral})^0$. 
Next-to-leading order (NLO) contributions
are again given by two tree diagrams
which however are of order $(Q / \Lambda_{chiral})^1$. 
These NLO diagrams arising from 
${\cal L}_2$ define the nucleon-weak current 
(or pion) vertices which include the nucleon recoil
correction of order $\sim m_N^{-1}$. 
To next-to-next-to-leading order (NNLO),  
$(Q / \Lambda_{chiral} )^2$, 
we have both tree and one-pion-loop diagrams. 
The tree diagrams here involve vertices 
coming from ${\cal L}_3$,  
which contains three low energy constants (LEC).   
Two of them can be determined from 
$<\!r^2\!>$ and $<\!r_A^2\!>$, 
while the third LEC can be constrained  
by the Goldberger-Treiman discrepancy. 
The one-pion-loop diagrams effectively introduce  
a form factor at the nucleon-pion vertex. 
It should be stressed that there 
are {\it no} free parameters in this ChPT calculation;  
all the LEC's are given by $g_A$, 
$\kappa_p-\kappa_n$, 
$f_\pi \simeq 93 $ MeV, $g_{\pi N}$, 
$<\!r^2\!>$, $<\!r^2_A\!>$, and the 
Goldberger-Treiman discrepancy $\Delta_\pi$
defined by $m_N g_A = f_\pi g_{\pi N} (1 - \Delta_\pi)$.  
Bernard {\it et al.}~\cite{bhm01} 
%,bernard-axial-review} 
have shown the very rapid convergence 
of these ChPT calculations for the spin-averaged 
OMC rate:
%\be
$\Lambda = ( 247 - 62 - 4 + \cdots ) {\rm s}^{-1}$,
%\ee
where the the first, second and third terms 
correspond to the LO, NLO and NNLO 
contributions, respectively. 
We can see that the NLO term is 
$\sim$25\% of the LO term 
and that the NNLO term is $\sim$1.6\% 
of the LO term. 
Meanwhile, the OMC rate from the 
$\mu$-$p$ singlet atomic state exhibits
a more subtle convergence behavior,
indicating that a systematic calculation
based on well-defined chiral expansion 
is indeed needed to get accurate predictions: 
\be
\Lambda_s ({\rm s}^{-1})= 957 - \frac{245 {\rm GeV}}{m_N} 
+ \left[ \frac{30.4 {\rm GeV}^2}{m_N^2} - 43.17\right] 
+ \cdots  \simeq 687 {\rm s}^{-1} \; . 
\ee
The second term in this expression is 
the nucleon recoil correction. 
Of the two terms in the square brackets
representing the NNLO corrections, 
the first represents the $m_N^{-2}$ correction term 
while the second term originates from 
the $q^2$-dependence in the hadronic form factors 
given by the ChPT terms of one-loop order. 
It is noteworthy that these two terms 
cancel each other to a significant degree,
a feature that cannot be studied reliably 
without a systematic expansion scheme
such as ChPT. 
We compare in Table \ref{table;OMCrate}
the OMC rates obtained in two independent ChPT calculations,
BHM~\cite{bhm01} and AMK~\cite{amk01}. 
The variance in the results of BHM and AMK
are ascribable to the slight differences 
in the approaches used.

\begin{table}
\begin{center}
%\begin{tabular}{|c|ccccc|} \hline
%  & This work & This work(NLO)&Bernard {\it et al.}
%\cite{bhm01} &
%Primakoff \cite{primakoff} & Opat \cite{opat}  \\ \hline
%$\Lambda_s $ & 695  & 722  & 711     & 664$\pm$20      & 634  \\
%$\Lambda_t $ & 11.9 & 12.2 & 14.0    & 11.9$\pm$0.7    & 13.3 
%\\ \hline
%\end{tabular} 
\begin{tabular}{|c|cccc|} \hline
  & BHM (NLO) & AMK (NLO)& BHM (NNLO)& AMK (NNLO)
\\ \hline
$\Lambda^{OMC}_s$ & 711  & 722  & 687     & 695  \\
$\Lambda^{OMC}_t$ & 14   & 12   & 13       & 12 
\\ \hline
\end{tabular} 
\caption{Comparison of calculated atomic OMC rates.  
$\Lambda_s$ ($\Lambda_t$) 
is the atomic capture rate (in sec$^{-1}$)
calculated for the initial singlet (triplet) 
hyperfine state including 
terms up to NLO or NNLO, as indicated.
The entries for the columns labeled ``BHM" 
and ``AMK" are taken from \cite{bhm01} and \cite{amk01}, 
respectively. 
The ``AMK" results have been obtained with the use of 
$g_A$ = 1.267 and $g_{\pi N}$ = 13.4. 
Apart from small chiral corrections,
the numerical results of the classic works by Primakoff and Opat 
would be close to those of AMK(NNLO), 
if the updated value of $g_A=1.267$ is adopted; 
Primakoff and Opat used $g_A = 1.24$ and $g_A=1.22$,
respectively }. 
\label{table;OMCrate}
\end{center}
\end{table}

The RMC atomic capture rates have been calculated 
with the same ${\cal L}_{eff}$. 
The use of ${\cal L}_{eff}$ ensures 
that the photon coupling in our RMC calculation 
automatically satisfies gauge invariance. 
The leading order (LO) diagram 
representing the emission of a photon by a muon  
gives a dominant contribution. 
In LO we also have Feynman diagrams 
of the following types:
(i) a photon couples to 
an intermediate pion propagator; 
(ii) a photon couples to a pion-nucleon vertex;
(iii) a photon couples to a $W^-$-pion vertex.
In the Coulomb gauge,
there are five LO Feynman diagrams, 
ten NLO diagrams, and 
more than twenty NNLO diagrams 
including pion-loop diagrams~\cite{am98}. 

It has been pointed out \cite{amk01b}
that, with the use of the atomic RMC rates
calculated model-independently with the use of HBChPT,
it is extremely difficult to reproduce 
$R_\gamma^{exp}$ obtained in the TRIUMF experiment.
We describe below some salient features of 
this difficulty.

\section{Muonic states in liquid hydrogen}

\begin{figure}
\parbox{.45\textwidth}{
\epsfig{file=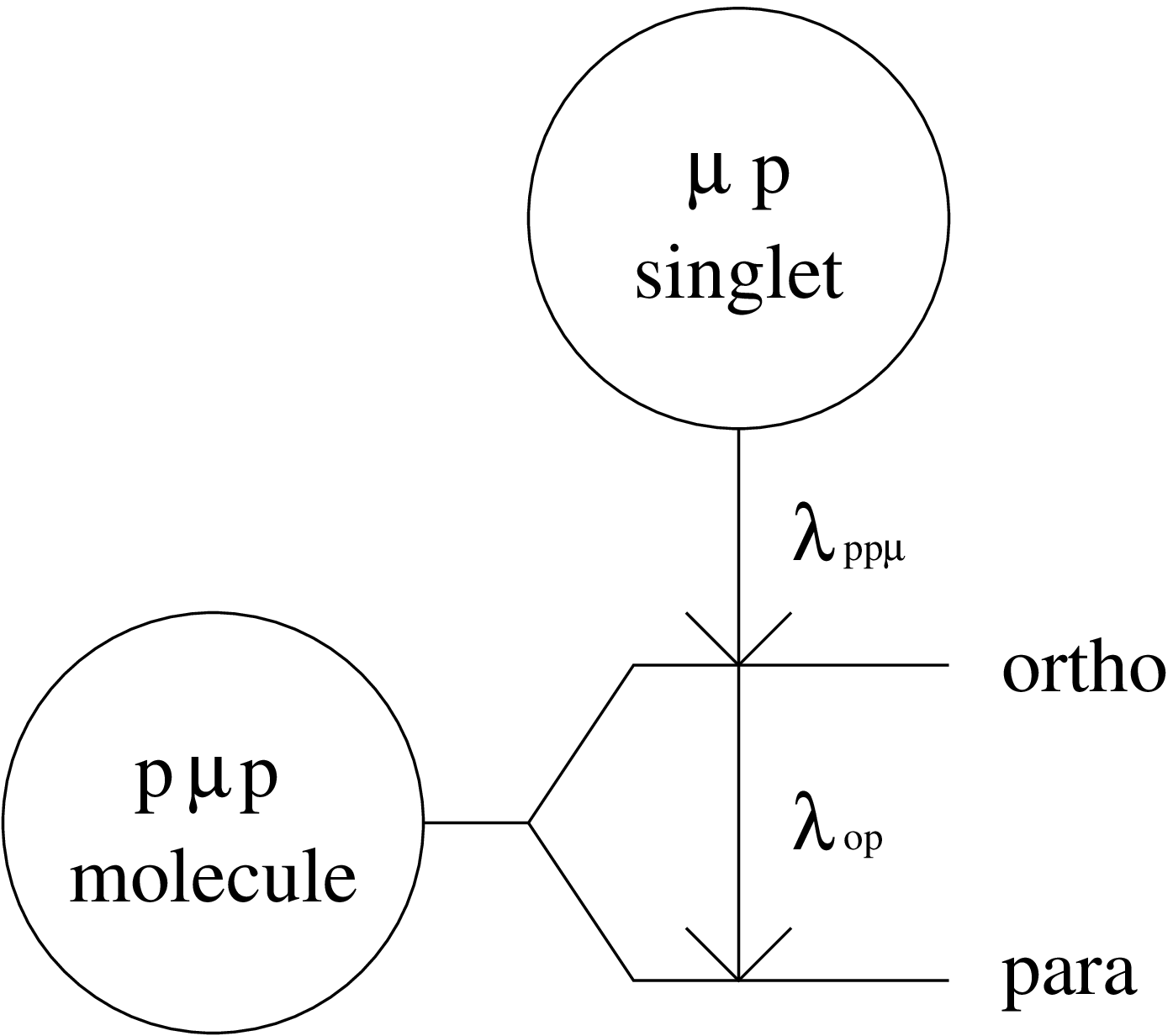,width=.3\textwidth,silent=,clip=}
\caption{\label{fig;liquid}
Atomic and molecular states relevant to 
muon capture in liquid hydrogen;
$\lambda_{pp\mu}$ is the transition rate 
from the atomic singlet state
to the ortho $p$-$\mu$-$p$ molecular state, 
and $\lambda_{op}$ is that
from the ortho to para molecular state.  }
}
\hfill
\parbox{.45\textwidth}{
\epsfig{file=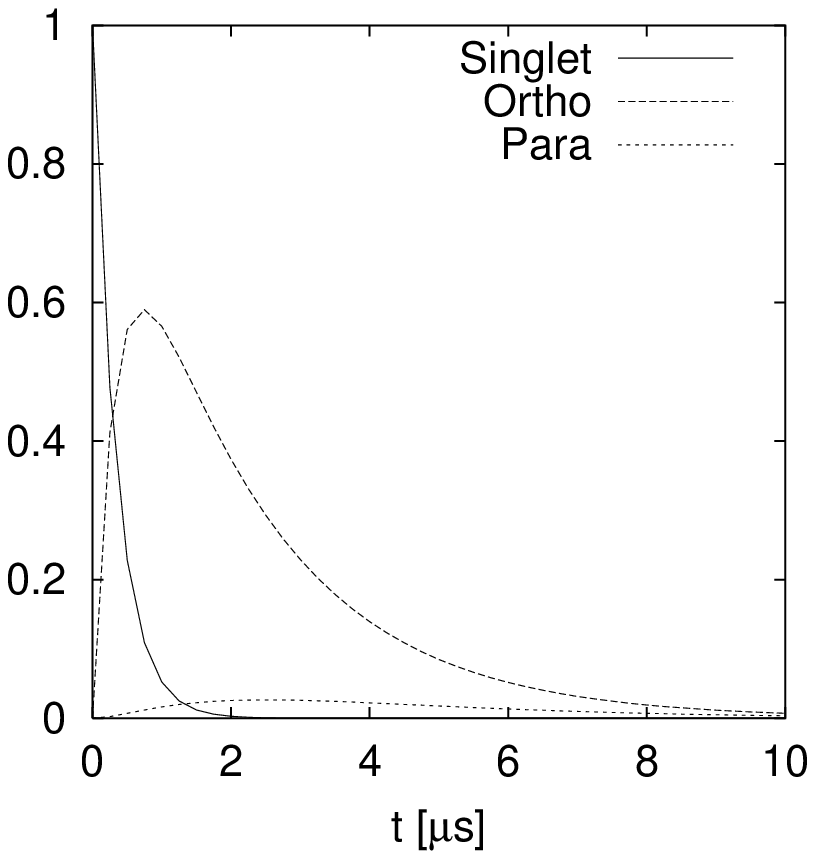,width=.43\textwidth,silent=,clip=}
\caption{\label{fig;states}
Number of muons at time $t$ in each state in liquid hydrogen.
} }
\end{figure}

To make a comparison between theory and experiment,
one needs to relate the theoretically calculated 
atomic OMC and RMC rates 
to the capture rates measured in a liquid H$_2$ target, 
$\Lambda_{liq}$ and $R_\gamma$, respectively. 
It is also important to know the temporal behavior of each of 
the various $\mu$-capture components
(capture from the atomic states
and capture from $p$-$\mu$-$p$ molecular states).
Fig.~\ref{fig;liquid} schematically
depicts various competing atomic and molecular  
processes in liquid $H_2$. 
A muon stopped in liquid hydrogen quickly forms 
a muonic atom ($\mu$-$p$) in the lowest Bohr state.
The atomic hyperfine-triplet state (S=1) 
decays extremely rapidly
to the singlet state (S=0),  
with a transition rate
$\lambda_{10}\simeq 1.7 \times 10^{10}$ s$^{-1}$.
In the liquid hydrogen target
a muonic atom and a hydrogen molecule 
collide with each other
to form a $p$-$\mu$-$p$ molecule 
with the molecule predominantly in its ortho state.
The transition rate from the atomic singlet state 
to the ortho $p$-$\mu$-$p$ molecular state, 
$\lambda_{pp\mu}$,  has an 
averaged value 
$\lambda_{pp\mu} \sim 2.5 \times 10^6$ s$^{-1}$,
which is comparable to the muon decay rate,  
$\lambda_0=0.455\times 10^6$ s$^{-1}$.  
The ortho $p$-$\mu$-$p$ state further decays to the para 
$p$-$\mu$-$p$ molecular state with a  
rate $\lambda_{op}$ $\sim (4 - 7) \times 10^4$ s$^{-1}$. 

%---------------------------------------------------------------
%-------------------------------------------------------------
%
%-------------------------------------------------------------------------
%\begin{figure}[t]
%\parbox{.55\textwidth}{\epsfig{file= 
%kltot.eps,width=.5\textwidth,silent=,clip=}}
%\hfill
%\parbox{.4\textwidth}{\caption{\label{sidebyside}
%Using \texttt{parbox}es, the figures may also be
%placed side-by-side with the caption. This may also be used
%to place several figures next to each other. (For an explanation of the
%figure, see Ref. \cite{fig2}.)}}
%\end{figure}
%------------------------------------------------------------------------

%We find that the 
%rates $\Lambda_{liq}$ and $R_\gamma$ are 
%sensitive to the value of $\lambda_{op}$ \cite{amk01b}.   
%Unfortunately
%there is a significant difference between
%the experimental and theoretical values;
%$\lambda_{op}^{exp}=
%(4.1\pm1.4)\times 10^4$ s$^{-1}$ \cite{bardin} 
%as compared with 
%$\lambda_{op}^{th} = (7.1\pm 1.2) \times 10^4$ s$^{-1}$ 
%\cite{bakalov}. 
%A TRIUMF experiment is making a precise $\lambda_{op}$ 
%measurement \cite{Lop-experiment}.

We now discuss very briefly the time structure 
relevant to the OMC experiment \cite{bardin}.
We denote by $N_s(t)$, $N_{om}(t)$ and $N_{pm}(t)$
the numbers of muons at time $t$ 
in the atomic singlet, ortho-molecular, 
and para-molecular states, respectively.
They satisfy coupled differential equations
(the kinetic equations), 
see Eq.~(54a) in Ref.~\cite{bakalov}. 
For illustration purposes, 
let us consider a case in which
there is one muon in the atomic singlet state at $t=0$;
{\it i.e.}, $N_s(0)=1$ and $N_{om}(0)=N_{pm}(0)=0$. 
$N_s(t)$, $N_{om}(t)$ and $N_{pm}(t)$
corresponding to this case are plotted
in Fig. \ref{fig;states}.
It turns out to be crucially important
to take proper account of the $t$-dependence
of these populations 
in analyzing the OMC data in \cite{bardin}, 
since data taking in Ref.~\cite{bardin} 
starts at $t\ne 0$. 

In the OMC experiment (see Fig.~4 in Ref.~\cite{bardin}),
$\mu^-$ beams arrive at the target  
in a (on the average) 3 $\mu$s-long burst
with repetition rate 3000 Hz.
The data collection typically starts 
1 $\mu$s after the end of the 3 $\mu$s-long beam burst,
and the measurement lasts until
306 $\mu$s after the end of the beam burst.  
To proceed, we assume 
that the quoted average time intervals
represent the actual values (ignoring fluctuations). 
Then, provided all the muons arrive at the same time,
we can choose with no ambiguity 
that arrival time as the origin of time ($t=0$)
and let $t$ = $t_i$, the starting time 
for data collection, 
refer to that origin. 
However, the finite duration (3 $\mu$s) 
of the beam burst causes uncertainty 
in the value of $t$ = $t_i$;   
$t_i$ can be anywhere 
between 1.0 $\mu$s and 4.0$ \mu$s. 
We choose here to average over 
the muon burst duration time 
for deducing $\Lambda_{liq}$, 
see Ref.~\cite{amk01b} for details.

For the RMC experiment 
\cite{jonkmans96,wright98},
the muons essentially arrive one by one 
and data taking begins at $t_i=365$ ns.
We therefore can neglect the beam burst duration time 
in calculating $R_\gamma$,
see Ref.~\cite{amk01b}. 

\section{Discussion}

The value of $\Lambda_{liq}$ obtained 
with the use of the atomic OMC rates calculated 
in HBChPT up to NNLO is  
$\Lambda_{liq}$ $\simeq$ 459 s$^{-1}$\cite{amk01,amk01b}.
This value is in good agreement with 
$\Lambda_{liq}^{exp}$. 
By contrast, it is not possible to reproduce
$R_\gamma^{exp}$ in Eq.(\ref{eq;exp-omc})  
in the existing theoretical framework
and with the use of the standard set of input
parameters, 
see Ref.~\cite{amk01b} for a more detailed discussion. 

The important question is:
Have we exhausted all possibile ways
for reconciling the measured OMC and RMC rates?
One thing worth studying
as a speculative possibility \cite{amk01b} 
is the sensitivity of 
$\Lambda_{liq}$ and $R_\gamma$ 
to a so-far neglected possible change in the value
of the molecular mixing parameter $\xi$.  
As first discussed by Weinberg~\cite{wei}, 
$\xi$ parametrizes a possible mixing
of the spin-3/2 and spin-1/2 states
in the $p$-$\mu$-$p$ molecule, and
this mixing changes the molecular capture rate to
\bea
\Lambda_{om}'= \xi 
\Lambda_{om}(1/2) + (1-\xi) \Lambda_{om}(3/2),
\eea
where $\Lambda_{om}(1/2) = \Lambda_{om}$ 
and $\Lambda_{om}(3/2) = 2\gamma_O\Lambda_t$ 
for both OMC and RMC;
$\Lambda_{om} = 2\gamma_O(0.75 \Lambda_s+0.25\Lambda_t)$
and $2\gamma_O=1.009$~\cite{bakalov}.
Although the existing theoretical estimate suggests
$\xi\simeq 1$~\cite{bakalov}, 
we treat $\xi$, as we did in Ref.~\cite{amk01},
as a parameter to fit the data.
Our study \cite{amk01b} demonstrates
that, even with this extra adjustable parameter,
it is impossible to fit 
the OMC and the RMC data simultaneously. 
In Ref.\cite{amk01b} 
we have also found that both the OMC and RMC capture rates are 
sensitive to the value of $\lambda_{op}$. 
Obviously, the results of a more precise measurement 
of $\lambda_{op}$ at TRIUMF \cite{Lop-experiment}
will be very important for both OMC and RMC. 
Note that since HBChPT constrains 
the value of $g_P$ with high accuracy,
there is not much room for adjusting 
the value of $g_P$. 

Our findings reported in \cite{amk01b}
and briefly summarized in this talk
are mostly the reconfirmation 
of the conclusions stated in one way 
or another in the literature,
but we hope that the coherent
treatment of OMC and RMC in liquid hydrogen 
as described in Ref.\cite{amk01b} would be useful.
Although we have presented examples
of simulation of the experimental conditions,
they are only meant to serve illustrative purposes.
Definitive analyses can be done only by the people
who carried out the relevant experiments.
Finally, we remark that
a precise measurement of the OMC rate in hydrogen gas
is planned at PSI\cite{PSI}. 
This experiment will allow us to  
avoid the molecular complexity  
discussed above and 
directly test the HBChPT prediction \cite{amk01,bhm01}.

\acknowledgments{
We are grateful to Peter Kammel, Tim Gorringe 
and Harold Fearing for useful comments.
This work is supported in part by NSF grant PHY-9900756.}

%%%%%%%%%%%%%%%%%%%%%%%%%%%%%%%%%%%%%%%%%%%%%%%%%%%%%%%%%%%%%%%%%%%%%%%%%%%%%%%%
%%
%____________________________________________________________
%  Start references here:

\end{document}